\documentclass[preprint, 3p, 11pt, times]{elsarticle}

\usepackage{amssymb}
\usepackage{amsmath}

\begin{document}

\begin{frontmatter}



\title{Human Mobility Reimagined: Digital Twin Intelligence for Adaptive Campus Course Timetabling}

\author[label1,label4]{Keshu Wu}
\ead{keshuw@tamu.edu}

\author[label2]{Xinyue Ye\corref{cor1}}
\ead{xye10@ua.edu}

\author[label1,label5]{Suphanut Jamonnak}
\ead{j.suphanut@tamu.edu}

\author[label3]{Xin Feng}
\ead{selena.feng@ou.edu}

\affiliation[label1]{organization={The Center for Geospatial Sciences, Applications, \& Technology, Department of Landscape Architecture and Urban Planning, Texas A\&M University},city={College Station}, state={TX}, country={USA}}

\affiliation[label4]{organization={Zachry Department of Civil and Environmental Engineering, Texas A\&M University},city={College Station}, state={TX}, country={USA}}

\affiliation[label2]{organization={Department of Geography \& the Environment, The University of Alabama},city={Tuscaloosa}, state={AL}, country={USA}}

\affiliation[label5]{organization={Texas A\&M Institute of Data Science },city={College Station}, state={TX}, country={USA}}

\affiliation[label3]{organization={Department of Geography \& Environmental Sustainability, The University of Oklahoma},city={Norman}, state={OK}, country={USA}}

\cortext[cor1]{Corresponding author.}

\begin{abstract}
Daily operations in large campuses depend on how efficiently people \emph{move} through space and time. In this sense, course timetables are more than administrative schedules: they act as mobility policies that orchestrate thousands of trajectories, shaping travel burden, congestion, accessibility, and the reliability of back-to-back transitions. Designing timetables that are both feasible and mobility-friendly is challenging because hard constraints including capacity, conflicts, feasibility must be satisfied alongside soft constraints including preferences, satisfaction, coordination, all under dynamic conditions such as real-time disruptions and evolving demand. Traditional static optimization methods often struggle to capture these human mobility impacts and to adapt when campus conditions change. This paper reconceptualizes course timetabling as a recommendation-based task and leverages the Texas A\&M Campus Digital Twin as a dynamic data platform to evaluate mobility consequences at scale. We propose an iterative framework that integrates collaborative and content-based filtering with feedback-driven refinement to generate ranked sets of adaptive timetable recommendations. A mobility-aware composite scoring function combining classroom occupancy, travel distance, travel time, and vertical transitions systematically balances resource efficiency with human-centered movement costs. Extensive experiments using real-world data from Texas A\&M University show that the proposed approach reduces mobility friction and travel inefficiencies, improves classroom utilization, and enhances overall user satisfaction. By coupling recommendation-oriented decision-making with digital twin intelligence, this study provides a robust and scalable blueprint for mobility-centered campus planning and resource allocation, with potential extensions to broader urban systems.

\end{abstract}


\end{frontmatter}

\section{Introduction}

Human mobility is a foundational layer of how complex systems function: it determines how people access services, how shared resources are utilized, and how daily activities unfold in space and time~\cite{barbosa2018human, gonzalez2008understanding, alessandretti2020scales, huang2025urban}. In increasingly data-rich environments, the central challenge is no longer merely to plan mobility under fixed assumptions, but to adapt mobility-centric decisions as conditions, preferences, and constraints evolve~\cite{pappalardo2023future, xugenerative}. Digital twins enable this shift by coupling a physical system with a continuously updated virtual counterpart, making it possible to evaluate decisions not only by feasibility and efficiency, but also by their mobility consequences~\cite{aghaabbasisustainability,ye2025toward,lin2024human,wang2022mobility}. This paper adopts that perspective and studies adaptive resource allocation through a human-mobility-first lens, using a university campus as a living microcosm where dense daily movements, tight time budgets, and heterogeneous indoor-outdoor infrastructure make mobility impacts immediately tangible.

Within this setting, a timetable is not simply an administrative artifact; it acts as a mobility policy that orchestrates thousands of day-to-day trajectories. Each assignment of an activity to a time and place implicitly shapes transition pressures, corridor congestion, and the uneven burdens of long walks, delayed arrivals, and vertical movements. Course timetabling in large academic institutions therefore represents a multifaceted challenge that requires the precise allocation of courses, classrooms, and time slots while meeting rigorous institutional policies and addressing the diverse needs of students and faculty \cite{chen2021survey, dinkel1989or, badri1998multi}. The problem is compounded by the necessity to balance \emph{hard constraints}—such as classroom capacities, scheduling conflicts, and feasible travel times—with \emph{soft constraints} including faculty preferences, student satisfaction, and inter-departmental coordination~\cite{hossain2019optimization, shiau2011hybrid}. Crucially, spatial-temporal factors such as building locations, pedestrian pathways, indoor circulation (e.g., corridors, stairs, elevators), and transition times introduce nontrivial dependencies between consecutive classes. These dependencies vary across individuals, time-of-day conditions, and infrastructure availability, making mobility inherently dynamic. Along with fluctuating enrollment, changing classroom availability, construction, and disruptions, these mobility considerations render traditional static optimization approaches insufficient \cite{burke2002recent, schaerf1999survey, lewis2008survey, hekmati2021course}. Consequently, there is an urgent need for adaptive methodologies that explicitly integrate spatial, temporal, and operational considerations to reduce mobility friction while maintaining institutional feasibility.

Extensive research has addressed the complexities of course timetabling using a variety of optimization techniques. Foundational approaches—such as constraint satisfaction models \cite{mccollum2006perspective} and combinatorial optimization frameworks \cite{qu2009hybridizations}—have laid the groundwork for automated timetabling. In addition, evolutionary algorithms, including genetic algorithms \cite{abdullah2008generating}, harmony search \cite{al2012harmony}, and particle swarm optimization \cite{tassopoulos2012solving}, have demonstrated the ability to navigate large solution spaces and balance competing objectives effectively. Mixed-integer programming (MIP) techniques provide exact solutions for smaller problems \cite{daskalaki2004integer, rappos2022mixed, al2007mixed}, while multi-objective optimization methods adeptly manage trade-offs between conflicting goals, such as minimizing travel distances while maximizing classroom utilization \cite{deb2011multi}. Moreover, hyper-heuristics have emerged as a flexible means to address diverse scheduling scenarios by balancing computational efficiency and solution quality \cite{qu2009hybridizations}. Despite these advances, many timetabling formulations still treat mobility as a coarse add-on (e.g., a single distance/time penalty) rather than modeling the full movement experience that includes inter-building flows, indoor-outdoor transitions, vertical movements, and time-varying congestion.

Recent advancements have further enhanced traditional methods by integrating machine learning and hybridization strategies. For instance, fuzzy logic has enabled the dynamic incorporation of user preferences into the scheduling process \cite{chaudhuri2010fuzzy}, and machine learning techniques have improved algorithm selection and design \cite{de2022algorithm, soria2014effective}. Clustering algorithms have been employed to simplify complex schedules \cite{kenekayoro2020incorporating}, and reinforcement learning has demonstrated promise in refining schedules through iterative interactions with dynamic environments \cite{yue2023reinforcement, obit2011non}. Nevertheless, significant gaps remain from a human mobility perspective. Many existing approaches do not fully incorporate spatial-temporal correlations in real campus networks or account for multi-modal transportation options (walking, bicycling, e-scooters, shuttles) and indoor mobility modes (corridors, stairs, elevators)—factors that directly determine travel feasibility, comfort, and schedule reliability in large academic settings. Additionally, most models lack the capability to generate adaptive schedules that respond in real time to disruptions and evolving requirements, even though mobility conditions and infrastructure availability can change within hours.

Digital twin technology has recently emerged as a transformative tool for addressing these challenges \cite{ye2023developing, ye2024developing}. A digital twin is a dynamic, real-time virtual representation of a physical system that seamlessly integrates spatial-temporal data with operational constraints, offering a powerful platform for simulation, monitoring, and optimization. While digital twins have been widely adopted in manufacturing to enhance production scheduling and operational efficiency \cite{tao2022digital, zhang2021digital, dasgupta2021transportation}, their application in urban systems and academic environments is still emerging. For campus operations, a digital twin provides the missing connective tissue between scheduling decisions and human mobility realities: it can encode building layouts, pathway networks, vertical circulation, and evolving conditions (e.g., closures or route changes), enabling timetables to be evaluated and refined based on realistic movement costs and constraints. Recent work in academic scheduling demonstrates the potential of digital twins to adapt dynamically to changes such as fluctuating enrollments, evolving classroom availability, and infrastructure disruptions \cite{ye2023developing}. By merging diverse datasets—including classroom attributes, transportation networks, and user feedback—digital twin technology enables a holistic approach to optimizing campus operations.

\begin{figure}[!t]
    \centering
    \includegraphics[width=\textwidth]{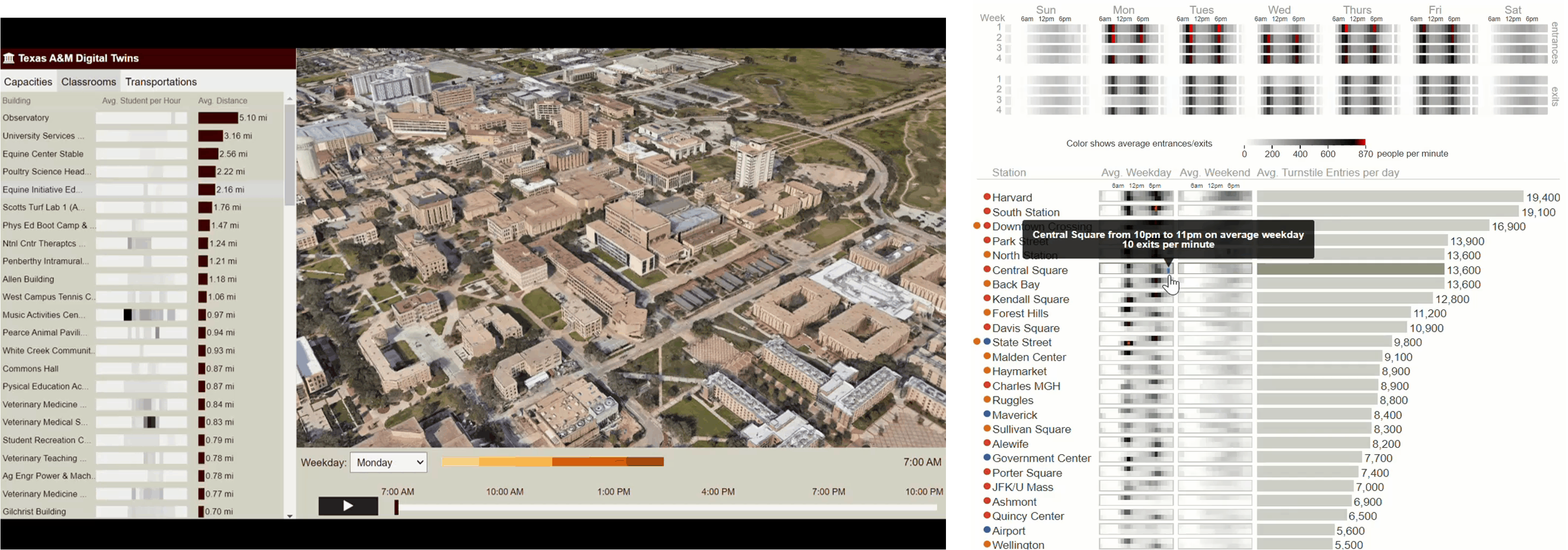}
    \caption{Texas A\&M Digital Twin.}
    \label{fig:tamu_dt}
\end{figure}

In this study, we leverage the Texas A\&M Campus Digital Twin \cite{ye2024developing, gong2025integrating}, as shown in Figure \ref{fig:tamu_dt}, to bridge the identified research gaps through a human-mobility-first perspective. We propose an iterative optimization framework for course timetabling that integrates spatial-temporal correlations, multi-modal transportation considerations, and user-centric mobility metrics. A composite scoring function evaluates course assignments based on normalized measures of classroom occupancy, travel distance, travel time, and vertical transitions between floors, thereby explicitly quantifying mobility burden alongside resource efficiency. The iterative optimization process then generates multiple feasible alternatives for bottleneck courses—those inducing excessive mobility costs or poor utilization—supporting targeted adjustments and dynamic responsiveness. By incorporating real-time feedback and leveraging digital twin capabilities, the proposed framework offers a robust and scalable solution for complex, dynamic scheduling requirements in large academic environments, while directly improving the lived mobility experience of campus users.

This research makes several key contributions. First, it reframes course scheduling as a human mobility-aware resource allocation problem and integrates spatial-temporal factors and multi-modal transportation options into the timetabling process, addressing a critical gap in existing methodologies. Second, it demonstrates the value of an iterative, feedback-driven optimization process for dynamically adapting to evolving constraints and mobility conditions. Third, it highlights the broader applicability of this approach beyond academic scheduling, extending to urban resource management and public facility planning where human movement and accessibility are central concerns. By emphasizing adaptability, transparency, and human-centric mobility outcomes, our work provides a comprehensive framework for addressing the complexities of modern timetabling.

The remainder of this paper is structured as follows: Section~\ref{sec:problem_statement} defines the problem settings and details the key variables and constraints. Section~\ref{sec:methodology} presents the proposed methodology, including data integration, scoring mechanisms, and the iterative optimization process. Section~\ref{sec:experiments} describes the experimental setup and evaluates the performance of the framework using real-world data from Texas A\&M University. Section~\ref{sec:discussion} discusses the implications of the findings, and Section~\ref{sec:conclusion} concludes with a summary of contributions and directions for future research.

\section{Problem Statement} \label{sec:problem_statement}

\subsection{Problem Settings}

Large-scale course timetabling at universities such as Texas A\&M University involves the systematic assignment of a set of courses $C = \{c_i \mid i = 1,\ldots,N\}$ to a network of classrooms $R = \{r_j \mid j = 1,\ldots,M\}$ across a series of time slots $T = \{t_k \mid k = 1,\ldots,K\}$ for a student population $V = \{v_l \mid l = 1,\ldots,L\}$. The objective is to construct a timetable that not only satisfies stringent institutional policies but also aligns with user preferences and optimizes resource utilization. The complexity arises from the need to meet both \emph{hard constraints} (e.g., classroom capacity, conflict avoidance, and travel feasibility) and \emph{soft constraints} (e.g., faculty preferences, degree requirements, and inter-departmental coordination). Moreover, spatial-temporal factors—such as building locations, travel routes, and transition times—introduce additional challenges. This multifaceted problem thus requires a comprehensive framework that assigns courses while simultaneously optimizing the ease of student movement across campus.

To address these challenges, we propose a robust and adaptive optimization framework that integrates multiple constraints into a unified composite objective function. This formulation is designed to balance strict feasibility requirements with user-centric quality measures, ensuring that the final timetable is both resource-optimal and conducive to smooth, efficient transitions.

\begin{figure}[!ht]
    \centering
    \includegraphics[width=\textwidth]{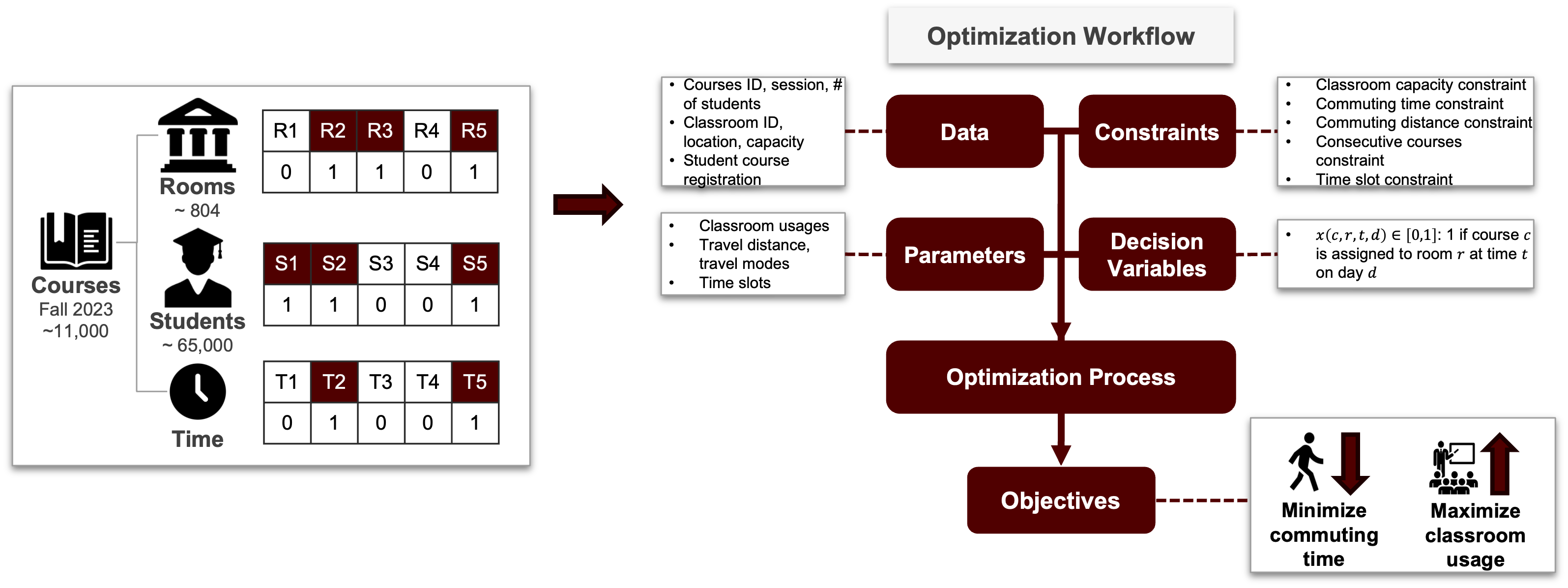}
    \caption{Overview of the Optimization Problem.}
    \label{fig:problem_overview}
\end{figure}

\subsection{Variables and Data Representation}

In our formulation, we adhere to the following index conventions:
\begin{itemize}
  \item \textbf{Courses:} \( c_i \) where \( i = 1,\ldots,N \).
  \item \textbf{Classrooms:} \( r_j \) where \( j = 1,\ldots,M \).
  \item \textbf{Time Slots:} \( t_k \) where \( k = 1,\ldots,K \).
  \item \textbf{Students:} \( v_l \) where \( l = 1,\ldots,L \).
\end{itemize}
These conventions ensure consistency and clarity throughout the model.

\paragraph{\textbf{Courses.}}  
Each course \(c_i\) is characterized by:
\begin{itemize}
    \item \textbf{Enrollment Set:} \( \phi(c_i) \subseteq V \), the set of students registered for course \(c_i\). The enrollment size is denoted by 
    \[
    E(c_i) = |\phi(c_i)|,
    \]
    which represents the number of students enrolled in \(c_i\). This metric is crucial for ensuring that room assignments meet capacity requirements.
    \item \textbf{Required Features:} \(F(c_i)\) denotes the specific facilities or equipment needed by \(c_i\) (e.g., AV equipment, specialized lab setups), ensuring that only compatible classrooms are considered.
    \item \textbf{Allowed Time Slots:} \(A(c_i) \subseteq T\) defines the set of time slots during which course \(c_i\) can be scheduled. This set reflects constraints arising from faculty availability and administrative policies.
\end{itemize}

\paragraph{\textbf{Classrooms}}  
Each classroom \(r_j\) is described by:
\begin{itemize}
    \item \textbf{Capacity:} \(\operatorname{Cap}(r_j)\) denotes the maximum number of students that room \(r_j\) can accommodate. This parameter is fundamental in ensuring that a room can physically host a given course.
    \item \textbf{Features:} \(F(r_j)\) provides a list of available resources in \(r_j\) (e.g., projectors, lab equipment) which are matched against \(F(c_i)\).
    \item \textbf{Location:} \(\operatorname{Loc}(r_j)\) specifies the spatial coordinates of \(r_j\), used in calculating travel metrics such as distance and travel time.
    \item \textbf{Transition History:} For each student \(v_l \in V\) and time slot \(t_k\), we define 
    \[
    r^{\text{prev}}(v_l,t_k),
    \]
    which represents the classroom where \(v_l\) attended a course immediately prior to \(t_k\). Formally, 
    \begin{equation}
        r^{\text{prev}}(v_l,t_k) =
        \begin{cases}
            r_{j'} & \text{if } \exists\, c_{i'} \text{ and } t_{k-1} \text{ such that } x_{i'j'(k-1)} = 1 \text{ and } v_l \in \phi(c_{i'}), \\
            \varnothing & \text{otherwise,}
        \end{cases}
    \end{equation}
    where \(x_{i'j'(k-1)} = 1\) indicates that course \(c_{i'}\) is scheduled in classroom \(r_{j'}\) at time \(t_{k-1}\); \(\varnothing\) indicates no prior assignment.
\end{itemize}

\paragraph{\textbf{Time Slots}}  
The set of time slots 
\[
T = \{t_1, t_2, \ldots, t_K\}
\]
represents discrete periods within the academic calendar, during which courses can be scheduled.

\paragraph{\textbf{spatial-temporal Data}}  
Additional campus-specific data is used to quantify transitions between classrooms:
\begin{itemize}
    \item \textbf{Distance Function:} \(d(r_j, r_{j'})\) represents the travel distance between classrooms \(r_j\) and \(r_{j'}\). This metric, derived from campus maps, quantifies spatial separation.
    \item \textbf{Time Function:} \(\tau(r_j, r_{j'})\) estimates the travel time between \(r_j\) and \(r_{j'}\). It reflects factors such as campus layout and pedestrian pathways.
    \item \textbf{Floor Function:} \(f(r_j, r_{j'})\) quantifies the number of floor transitions (vertical movements) required between \(r_j\) and \(r_{j'}\), capturing the physical effort required.
\end{itemize}

\paragraph{\textbf{Decision Variable}}  
The binary decision variable is defined as:
\[
x_{ijk} =
\begin{cases}
1, & \text{if course } c_i \text{ is assigned to classroom } r_j \text{ at time } t_k, \\
0, & \text{otherwise.}
\end{cases}
\]
This variable is central to the formulation, linking courses, classrooms, and time slots.

\paragraph{\textbf{Auxiliary Variables for Transition Metrics}}  
For each assignment \((c_i, r_j, t_k)\) and each student \(v_l \in \phi(c_i)\), we define:
\[
\delta_{ijk}^{(l),d} = d\Bigl(r^{\text{prev}}(v_l,t_k),\, r_j\Bigr),
\]
\[
\delta_{ijk}^{(l),\tau} = \tau\Bigl(r^{\text{prev}}(v_l,t_k),\, r_j\Bigr),
\]
\[
\delta_{ijk}^{(l),f} = f\Bigl(r^{\text{prev}}(v_l,t_k),\, r_j\Bigr).
\]
Here, \(\delta_{ijk}^{(l),d}\) is the travel distance, \(\delta_{ijk}^{(l),\tau}\) is the travel time, and \(\delta_{ijk}^{(l),f}\) is the number of floor transitions for student \(v_l\) when transitioning to classroom \(r_j\) at time \(t_k\). If \(r^{\text{prev}}(v_l,t_k) = \varnothing\), these values are set to zero (or another nominal value) to indicate no transition penalty.

\subsection{Objective Function}

The objective is to maximize the overall quality of the timetable by balancing classroom utilization and the ease of student transitions. To this end, we construct a composite scoring function that integrates four normalized metrics—occupancy, travel distance, travel time, and floor transitions—each evaluated for a specific assignment \((c_i, r_j, t_k)\). Each metric is normalized to \([0, 1]\) and weighted by a corresponding coefficient reflecting institutional priorities.

\paragraph{\textbf{Occupancy Score (\(S_O\))}}  
For an assignment of course \(c_i\) to classroom \(r_j\), the occupancy score is defined as:
\begin{equation}
S_O(c_i, r_j) = \frac{E(c_i)}{\operatorname{Cap}(r_j)},
\end{equation}
where \(E(c_i)\) is the number of students enrolled in \(c_i\) and \(\operatorname{Cap}(r_j)\) is the capacity of \(r_j\). A value near 1 indicates that the classroom is being used efficiently. We compute the overall average occupancy score as:
\begin{equation}
\bar{S}_O = \frac{\sum_{i=1}^{N} \sum_{j=1}^{M} \sum_{k \in A(c_i)} x_{ijk}\, S_O(c_i, r_j)}{\sum_{i=1}^{N} \sum_{j=1}^{M} \sum_{k \in A(c_i)} x_{ijk}},
\end{equation}
where the numerator aggregates the occupancy scores over all active assignments (i.e., where \(x_{ijk}=1\)) and the denominator counts the total number of assignments.

\paragraph{\textbf{Travel Distance Score (\(S_D\))}}  
For each student \(v_l \in \phi(c_i)\) associated with an assignment \((c_i, r_j, t_k)\), the individual travel distance score is:
\[
s_{ijk}^{(l),d} = 1 - \frac{\delta_{ijk}^{(l),d}}{\Delta_{d}},
\]
where \(\Delta_{d}\) is a constant representing the maximum acceptable travel distance. The assignment-level travel distance score is then given by:
\begin{equation}
S_D(c_i, r_j, t_k) = \frac{1}{E(c_i)} \sum_{v_l \in \phi(c_i)} s_{ijk}^{(l),d}.
\end{equation}
The overall average travel distance score is:
\begin{equation}
\bar{S}_D = \frac{\sum_{i=1}^{N} \sum_{j=1}^{M} \sum_{k \in A(c_i)} x_{ijk}\, S_D(c_i, r_j, t_k)}{\sum_{i=1}^{N} \sum_{j=1}^{M} \sum_{k \in A(c_i)} x_{ijk}}.
\end{equation}
This metric captures the spatial feasibility of transitions, favoring assignments that minimize overall travel distances for students.

\paragraph{\textbf{Travel Time Score (\(S_T\))}}  
Analogously, for each student \(v_l \in \phi(c_i)\), the individual travel time score is:
\[
s_{ijk}^{(l),\tau} = 1 - \frac{\delta_{ijk}^{(l),\tau}}{\Delta_{\tau}},
\]
where \(\Delta_{\tau}\) is the maximum acceptable travel time. The assignment-level travel time score is:
\begin{equation}
S_T(c_i, r_j, t_k) = \frac{1}{E(c_i)} \sum_{v_l \in \phi(c_i)} s_{ijk}^{(l),\tau},
\end{equation}
and the overall average travel time score is:
\begin{equation}
\bar{S}_T = \frac{\sum_{i=1}^{N} \sum_{j=1}^{M} \sum_{k \in A(c_i)} x_{ijk}\, S_T(c_i, r_j, t_k)}{\sum_{i=1}^{N} \sum_{j=1}^{M} \sum_{k \in A(c_i)} x_{ijk}}.
\end{equation}
This score ensures that assignments with shorter transition times are preferred, which is particularly beneficial during tight scheduling.

\paragraph{\textbf{Travel Floor Score (\(S_F\))}}  
For each student \(v_l \in \phi(c_i)\), the individual floor score is defined as:
\[
s_{ijk}^{(l),f} = 1 - \frac{\delta_{ijk}^{(l),f}}{\Delta_{f}},
\]
where \(\Delta_{f}\) represents the maximum allowed number of floor transitions. The assignment-level floor score is:
\begin{equation}
S_F(c_i, r_j, t_k) = \frac{1}{E(c_i)} \sum_{v_l \in \phi(c_i)} s_{ijk}^{(l),f},
\end{equation}
and the overall average floor score is:
\begin{equation}
\bar{S}_F = \frac{\sum_{i=1}^{N} \sum_{j=1}^{M} \sum_{k \in A(c_i)} x_{ijk}\, S_F(c_i, r_j, t_k)}{\sum_{i=1}^{N} \sum_{j=1}^{M} \sum_{k \in A(c_i)} x_{ijk}}.
\end{equation}
This metric quantifies the physical burden of vertical transitions and promotes assignments that minimize such inconveniences.

\paragraph{\textbf{Composite Objective}}  
The overall objective is to maximize the weighted average of these aggregated scores:
\begin{equation}\label{eq:scoring_function}
\text{Maximize } Z = \alpha_1 \bar{S}_O + \alpha_2 \bar{S}_D + \alpha_3 \bar{S}_T + \alpha_4 \bar{S}_F,
\end{equation}
where \(\alpha_1, \alpha_2, \alpha_3, \alpha_4 \geq 0\) are weighting coefficients determined via sensitivity analysis. This composite objective ensures that the final timetable achieves a balance between efficient resource utilization and the minimization of student transition burdens.

\subsection{Hard Constraints}

The feasibility of the timetable is enforced through hard constraints that are directly linked to the decision variable \(x_{ijk}\).

\paragraph{\textbf{Capacity Constraint}}  
If course \(c_i\) is assigned to classroom \(r_j\) at time \(t_k\) (i.e., \(x_{ijk} = 1\)), then the classroom must have sufficient capacity to accommodate all enrolled students:
\begin{equation}
x_{ijk} \cdot E(c_i) \leq \operatorname{Cap}(r_j), \quad \forall\, i=1,\ldots,N,\; j=1,\ldots,M,\; k=1,\ldots,K.
\end{equation}

\paragraph{\textbf{Travel Time Feasibility}}  
For every active assignment \((c_i, r_j, t_k)\) with \(k > 1\) and for every student \(v_l \in \phi(c_i)\), the travel time from the previous classroom must not exceed the allowable limit:
\begin{equation}
x_{ijk} \cdot \tau\Bigl(r^{\text{prev}}(v_l,t_k), r_j\Bigr) \leq \Delta_{\tau}, \quad \forall\, i=1,\ldots,N,\; j=1,\ldots,M,\; k=2,\ldots,K,\; \forall\, v_l \in \phi(c_i).
\end{equation}

\paragraph{\textbf{Travel Distance Feasibility}}  
Similarly, for each active assignment and for every student \(v_l \in \phi(c_i)\), the travel distance must not exceed the prescribed threshold:
\begin{equation}
x_{ijk} \cdot d\Bigl(r^{\text{prev}}(v_l,t_k), r_j\Bigr) \leq \Delta_{d}, \quad \forall\, i=1,\ldots,N,\; j=1,\ldots,M,\; k=2,\ldots,K,\; \forall\, v_l \in \phi(c_i).
\end{equation}

\paragraph{\textbf{Travel Floor Feasibility}}  
For vertical transitions, the number of floor changes must remain within the allowed limit:
\begin{equation}
x_{ijk} \cdot f\Bigl(r^{\text{prev}}(v_l,t_k), r_j\Bigr) \leq \Delta_{f}, \quad \forall\, i=1,\ldots,N,\; j=1,\ldots,M,\; k=2,\ldots,K,\; \forall\, v_l \in \phi(c_i).
\end{equation}

\paragraph{\textbf{Conflict Avoidance}}  
To ensure that every course is scheduled exactly once, we impose:
\begin{equation}
\sum_{j=1}^{M} \sum_{k \in A(c_i)} x_{ijk} = 1, \quad \forall\, i=1,\ldots,N.
\end{equation}
This constraint guarantees that each course \(c_i\) is assigned to one and only one classroom and one time slot, thereby eliminating scheduling conflicts.

\subsection{Soft Constraints}

While hard constraints guarantee the feasibility of the timetable, soft constraints enhance its overall quality by introducing flexibility. Examples include:
\begin{itemize}
    \item \textbf{Degree Requirements:} Courses critical for degree completion are prioritized to avoid conflicts that might impede student progress.
    \item \textbf{Faculty Preferences:} Incorporation of faculty availability and preferred teaching times/locations improves overall satisfaction.
    \item \textbf{Cross-Departmental Coordination:} Minimizing scheduling conflicts among courses shared by multiple departments enhances inter-departmental cooperation.
    \item \textbf{Student Course Clustering:} Grouping courses frequently taken together in close temporal and spatial proximity reduces transition burdens.
    \item \textbf{Resource Matching:} Aligning classroom assignments with course-specific resource needs (e.g., specialized laboratories) optimizes facility utilization.
\end{itemize}
These soft constraints are embedded within the composite scoring function (e.g., via \(S_O\), \(S_D\), \(S_T\), and \(S_F\)) and may be further augmented by additional user preference metrics such as \(P(c_i,t_k)\).

\subsection{Complexity and Proposed Solution}

The integration of numerous hard and soft constraints renders the course timetabling problem a high-dimensional, combinatorial challenge. Traditional static optimization methods often struggle to capture the dynamic, user-centric aspects inherent in such problems. To address these challenges, we propose an iterative recommendation framework that incorporates real-time feedback, detailed spatial-temporal analysis, and dynamic scoring adjustments. This adaptive approach is designed to converge to a high-quality timetable that effectively harmonizes institutional objectives with user satisfaction, thus achieving enhanced adaptability, robustness, and operational efficiency.

\subsection{Notations and Definitions}

Table~\ref{tab:notations} summarizes the main notations used throughout this paper along with their respective meanings.

\begin{table}[!ht]
\scriptsize
\centering
\caption{Summary of Notations and Definitions}
\label{tab:notations}
\begin{tabular}{c|p{9cm}}
\textbf{Notation} & \textbf{Definition} \\ 
\hline
\( C = \{ c_i \mid i=1,\ldots,N \} \) & Set of courses, where \( c_i \) denotes the \(i\)th course. \\ 
\( R = \{ r_j \mid j=1,\ldots,M \} \) & Set of classrooms, where \( r_j \) denotes the \(j\)th classroom. \\ 
\( T = \{ t_k \mid k=1,\ldots,K \} \) & Set of time slots, where \( t_k \) denotes the \(k\)th time slot. \\ 
\( V = \{ v_l \mid l=1,\ldots,L \} \) & Set of students, where \( v_l \) denotes the \(l\)th student. \\ 
\( \phi(c_i) \) & Enrollment set for course \( c_i \); i.e., the set of students registered for \( c_i \). \\ 
\( E(c_i) \) & Enrollment size of course \( c_i \), representing the number of students enrolled. \\ 
\( F(c_i) \) & Required features for course \( c_i \) (e.g., AV equipment, lab setups). \\ 
\( A(c_i) \subseteq T \) & Allowed time slots for course \( c_i \). \\ 
\( \operatorname{Cap}(r_j) \) & Capacity of classroom \( r_j \) (maximum number of students it can accommodate). \\ 
\( F(r_j) \) & Features available in classroom \( r_j \) (e.g., projectors, lab equipment). \\ 
\( \operatorname{Loc}(r_j) \) & Spatial coordinates of classroom \( r_j \). \\ 
\( r^{\text{prev}}(v_l,t_k) \) & Previous classroom attended by student \( v_l \) before time \( t_k \). \\ 
\( d(r_j, r_{j'}) \) & Travel distance between classrooms \( r_j \) and \( r_{j'} \). \\ 
\( \tau(r_j, r_{j'}) \) & Estimated travel time between classrooms \( r_j \) and \( r_{j'} \). \\ 
\( f(r_j, r_{j'}) \) & Number of floor transitions between classrooms \( r_j \) and \( r_{j'} \). \\ 
\( x_{ijk} \) & Binary decision variable: \( x_{ijk} = 1 \) if course \( c_i \) is assigned to classroom \( r_j \) at time \( t_k \), and 0 otherwise. \\ 
\( \delta_{ijk}^{(l),d} \) & Travel distance for student \( v_l \) from \( r^{\text{prev}}(v_l,t_k) \) to \( r_j \) at time \( t_k \). \\ 
\( \delta_{ijk}^{(l),\tau} \) & Travel time for student \( v_l \) from \( r^{\text{prev}}(v_l,t_k) \) to \( r_j \) at time \( t_k \). \\ 
\( \delta_{ijk}^{(l),f} \) & Number of floor transitions for student \( v_l \) from \( r^{\text{prev}}(v_l,t_k) \) to \( r_j \) at time \( t_k \). \\ 
\( S_O(c_i, r_j) \) & Occupancy score for assignment of course \( c_i \) to classroom \( r_j \). \\ 
\( S_D(c_i, r_j, t_k) \) & Travel distance score for assignment \((c_i, r_j, t_k)\). \\ 
\( S_T(c_i, r_j, t_k) \) & Travel time score for assignment \((c_i, r_j, t_k)\). \\ 
\( S_F(c_i, r_j, t_k) \) & Travel floor score for assignment \((c_i, r_j, t_k)\). \\ 
\( \bar{S}_O, \bar{S}_D, \bar{S}_T, \bar{S}_F \) & Overall average scores for occupancy, travel distance, travel time, and travel floor, respectively, aggregated over all active assignments. \\ 
\( \alpha_1, \alpha_2, \alpha_3, \alpha_4 \) & Weighting coefficients for occupancy, travel distance, travel time, and travel floor scores, respectively, determined via sensitivity analysis. \\ 
\( Z \) & Composite objective value to be maximized. \\ 
\( p^{\text{out}}_m \) & Probability of using outdoor transportation mode \(m\) (e.g., walking, bicycling). \\ 
\( p^{\text{in}}_m \) & Probability of using indoor transportation mode \(m\) (e.g., stairs, elevators). \\ 
\( \tau^{\text{out}}_{\text{eff}}(r_j, r_{j'}) \) & Effective outdoor travel time between classrooms \(r_j\) and \(r_{j'}\). \\ 
\( \tau^{\text{in}}_{\text{eff}}(r_j, r_j^{exit}) \) & Effective indoor travel time from classroom \(r_j\) to its building's exit point \(r_j^{exit}\). \\ 
\( \tau^{\text{in}}_{\text{eff}}(r_{j'}^{entry}, r_{j'}) \) & Effective indoor travel time from the entry point \(r_{j'}^{entry}\) of the building containing \(r_{j'}\) to \(r_{j'}\). \\ 
\end{tabular}
\end{table}

\section{Methodology} \label{sec:methodology}

\subsection{Conceptual Overview}

The proposed recommendation system builds upon the Texas A\&M Campus Digital Twin, which serves as the foundational data platform. This digital twin seamlessly integrates building layouts, classroom attributes, and real-time operational data—including updates on transportation routes and building availability—to create an accurate and dynamic representation of campus conditions. The methodology is specifically designed to tackle the complex and dynamic nature of course timetabling by incorporating both hard constraints (e.g., capacity limits, conflict avoidance, and travel feasibility) and soft constraints (e.g., faculty preferences and degree requirements). Furthermore, it optimizes resource allocation and enhances user satisfaction by considering spatial and temporal dependencies.

The overall framework comprises three key layers:

\begin{enumerate}
    \item \textbf{Data Integration and Preprocessing:} This layer consolidates heterogeneous data sources into a unified schema, ensuring consistency and reliability. It integrates information such as building capacities, classroom features, course requirements, student schedules, and real-time environmental conditions.
    \item \textbf{Recommendation Engine:} At the heart of the system, the recommendation engine employs a hybrid approach that combines collaborative filtering with content-based filtering to generate personalized and spatial-temporal recommendations. This engine leverages both historical data and real-time updates to propose optimal timetabling solutions.
    \item \textbf{Iterative Adjustment and Feedback:} This layer dynamically refines the generated recommendations based on continuous user feedback and real-time data updates. By iteratively adjusting the recommendations, the system maintains alignment with institutional goals and evolving user needs.
\end{enumerate}

\subsection{Data Representation}

To effectively manage the multi-dimensional nature of the problem, the system uses three core datasets: (i) building and classroom information, (ii) course information, and (iii) student information. These datasets provide the necessary inputs for the scoring and optimization process, ensuring seamless integration and adaptability.

\subsubsection{Building and Classroom Information}

Each building is uniquely identified and represented by its spatial coordinates and a nested structure containing classroom details. This hierarchical data structure is essential for evaluating both feasibility and optimization criteria. Table~\ref{tab:building} illustrates the building-level data structure, while Table~\ref{tab:classroom} provides details on classroom-specific attributes.

\begin{table}[!ht]
\scriptsize
\centering
\caption{Building and Classroom Data Structure}
\label{tab:building}
\begin{tabular}{l|l|l}

\textbf{Field}         & \textbf{Description}                                        & \textbf{Example}            \\ \hline
\texttt{building\_code} & Unique identifier for the building.                         & \texttt{ARCC}               \\ 
\texttt{building\_name} & Full name of the building.                                  & Architecture Center, Bldg C \\ 
\texttt{latitude}       & Latitude coordinate for spatial calculations.               & 30.615093                   \\ 
\texttt{longitude}      & Longitude coordinate for spatial calculations.              & -96.341195                  \\ 
\texttt{classrooms}     & Nested list containing classroom-specific details.          & See Table~\ref{tab:classroom}. \\ 
\end{tabular}
\end{table}

\begin{table}[!ht]
\scriptsize
\centering
\caption{Classroom-Specific Details}
\label{tab:classroom}
\begin{tabular}{l|l|l}

\textbf{Field}      & \textbf{Description}                     & \textbf{Example} \\ \hline
\texttt{room\_number} & Room number in the building.             & 101              \\ 
\texttt{capacity}    & Maximum student capacity.                & 50               \\ 
\texttt{room\_timetable} & List of scheduled courses and timings.  & \{...\}, \{...\} \\ 
\end{tabular}
\end{table}

\subsubsection{Course Information}
Each course entry includes a variety of scheduling details, room requirements, and enrollment data. These data elements ensure that courses are scheduled in a manner that respects both capacity and spatial-temporal constraints. Table~\ref{tab:course} outlines the structure of the course dataset.

\begin{table}[!ht]
\scriptsize
\centering
\caption{Course Data Structure}
\label{tab:course}
\begin{tabular}{l|l|l}

\textbf{Field}            & \textbf{Description}                                         & \textbf{Example} \\ \hline
\texttt{course\_id}       & Unique identifier for the course.                            & COSC101-001      \\ 
\texttt{meeting\_id}      & Unique meeting identifier.                                   & M001             \\ 
\texttt{type}             & Type of meeting (e.g., lecture, lab).                        & LEC              \\ 
\texttt{days}             & Days of the week the course meets.                           & MWF              \\ 
\texttt{start\_time}      & Start time of the course meeting.                            & 09:00            \\ 
\texttt{end\_time}        & End time of the course meeting.                              & 09:50            \\ 
\texttt{building\_code}   & Code for the building where the course is held.              & ARCC             \\ 
\texttt{room\_number}     & Classroom number where the course meets.                     & 101              \\ 
\texttt{enrolled}         & Number of students enrolled.                                 & 45               \\ 
\texttt{capacity\_required} & Minimum capacity required for the course.                   & 50               \\ 
\end{tabular}
\end{table}

\subsubsection{Student Information}
Student data encompasses individual schedules, personal preferences, and enrollment records, facilitating personalized scheduling and the implementation of soft constraints such as clustering courses that are commonly taken together. Table~\ref{tab:student} details the student data structure, while Table~\ref{tab:schedule} describes the nested schedule format.

\begin{table}[!ht]
\scriptsize
\centering
\caption{Student Data Structure}
\label{tab:student}
\begin{tabular}{l|l|l}

\textbf{Field}         & \textbf{Description}                                         & \textbf{Example} \\ \hline
\texttt{student\_id}    & Unique identifier for the student.                          & S123456          \\ 
\texttt{courses}       & List of enrolled courses for the student.                    & \{COSC101-001\}  \\ 
\texttt{schedule}      & Details of the student's course schedules (nested).          & See Table~\ref{tab:schedule}. \\ 
\end{tabular}
\end{table}

\begin{table}[!ht]
\scriptsize
\centering
\caption{Schedule Nested Structure}
\label{tab:schedule}
\begin{tabular}{l|l|l}

\textbf{Field}        & \textbf{Description}                   & \textbf{Example} \\ \hline
\texttt{course\_id}   & Identifier of the enrolled course.     & COSC101-001      \\ 
\texttt{start\_time}  & Start time of the course.              & 09:00            \\ 
\texttt{end\_time}    & End time of the course.                & 09:50            \\ 
\texttt{building\_code} & Code for the building location.       & ARCC             \\ 
\texttt{room\_number} & Classroom number for the course.       & 101              \\ 
\end{tabular}
\end{table}

\subsection{System Highlights}

The proposed recommendation system leverages both spatial-temporal factors and multi-modal transportation data to produce high-quality timetabling solutions. Two key aspects of the system are highlighted below, each incorporating quantitative measures to enhance decision-making.

\paragraph{\textbf{Spatial-Temporal Integration}}  
By integrating detailed spatial features (such as building locations and route networks) and temporal features (such as course durations and commuting times), the system computes critical metrics including travel distance, travel time, and the number of floor transitions between classroom assignments. These metrics are derived from functions \(d(r_j, r_{j'})\) for distance, \(\tau(r_j, r_{j'})\) for travel time, and \(f(r_j, r_{j'})\) for floor transitions. These computations are central to enforcing hard constraints—such as ensuring that travel times remain below a 20-minute threshold and that classrooms are within walkable distances—thereby minimizing transit burdens for students.

\paragraph{\textbf{Multi-Modal Transportation Options}}  
To accurately capture real-world conditions, the system distinguishes between indoor and outdoor transportation modes. For outdoor transit, we consider modes such as walking, bicycling, e-scooters, and campus shuttles. Each outdoor mode is assigned a probability $p^{\text{out}}_m$, with the constraint that $\sum_{m \in \text{Outdoor}} p^{\text{out}}_m = 1$. Similarly, for indoor transit, where options include corridors, stairs, and elevators, we similarly assign probabilities $p^{\text{in}}_m$ such that $\sum_{m \in \text{Indoor}} p^{\text{in}}_m = 1$.

\begin{figure}[!ht]
    \centering
    \includegraphics[width=0.65\textwidth]{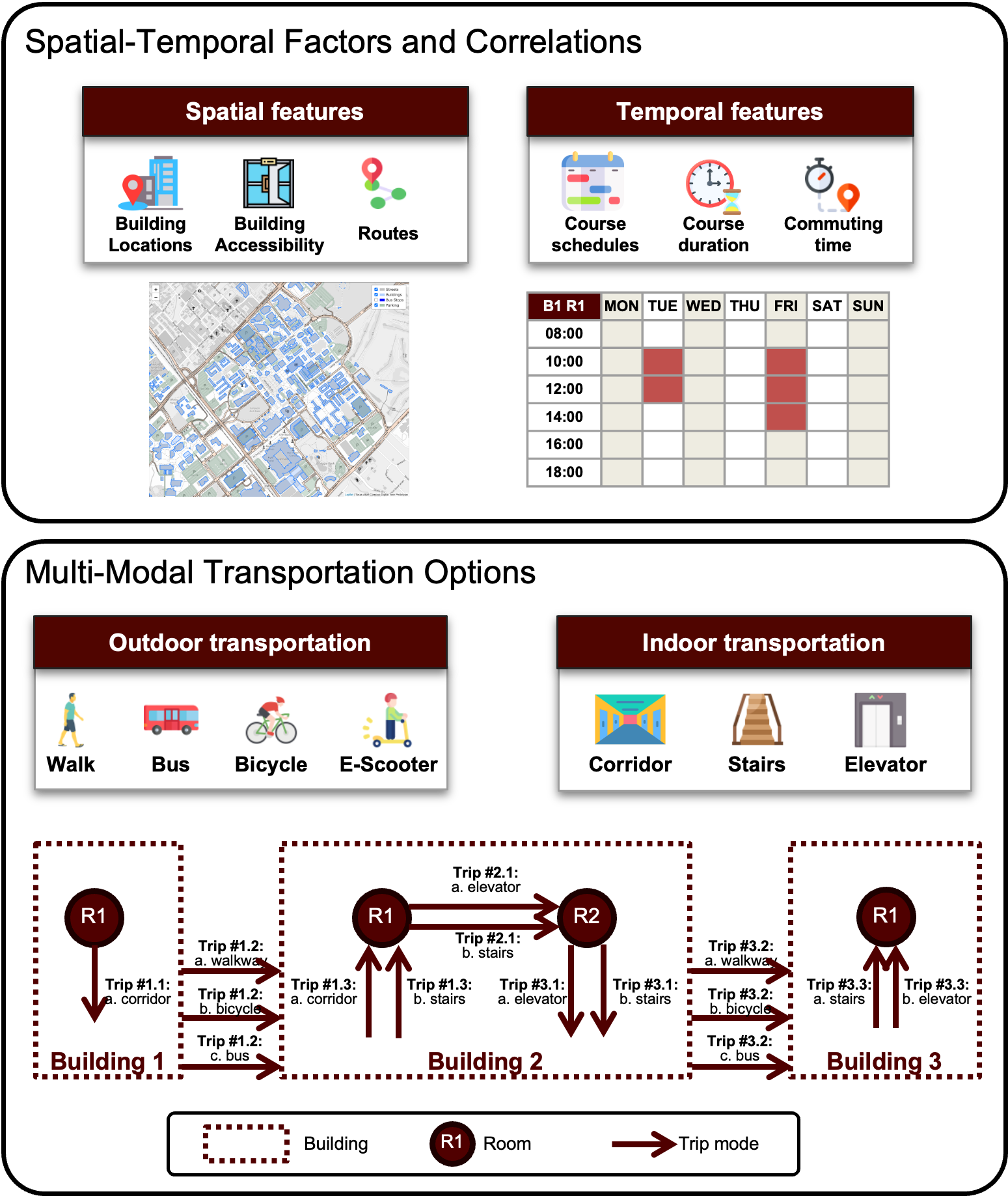}
    \caption{Highlight 1: Spatial-Temporal Factors and Correlations and Highlight 2: Multi-Modal Transportation Options.}
    \label{fig:highlights}
\end{figure}

The effective travel time between two classrooms is modeled using a conditional function that distinguishes between same-building and different-building scenarios. In cases where both classrooms are located in the same building, only one indoor transit is needed. However, when the classrooms belong to different buildings, students must complete two indoor trips (one for exiting the origin building and another for entering the destination building) in addition to the outdoor transit between buildings. This can be formalized as:
\begin{equation}
\tau_{\text{eff}}(r_j, r_{j'}) = 
\begin{cases} 
\tau^{\text{in}}_{\text{eff}}(r_j, r_{j'}) & \text{if } \operatorname{Building}(r_j) = \operatorname{Building}(r_{j'}), \\[2mm]
\tau^{\text{out}}_{\text{eff}}(r_j, r_{j'}) + \tau^{\text{in}}_{\text{eff}}(r_j, r_j^{exit}) + \tau^{\text{in}}_{\text{eff}}(r_{j'}^{entry}, r_{j'}) & \text{if } \operatorname{Building}(r_j) \neq \operatorname{Building}(r_{j'}),
\end{cases}
\end{equation}
where:
\begin{equation}
\tau^{\text{out}}_{\text{eff}}(r_j, r_{j'}) = \sum_{m \in \text{Outdoor}} p^{\text{out}}_m \, \tau_m(r_j, r_{j'}),
\end{equation}
\begin{equation}
\tau^{\text{in}}_{\text{eff}}(r_j, r_j^{exit}) = \sum_{m \in \text{Indoor}} p^{\text{in}}_m \, \tau_m(r_j, r_j^{exit}),
\end{equation}
\begin{equation}
\tau^{\text{in}}_{\text{eff}}(r_{j'}^{entry}, r_{j'}) = \sum_{m \in \text{Indoor}} p^{\text{in}}_m \, \tau_m(r_{j'}^{entry}, r_{j'}).
\end{equation}
Here, $\tau_m(r_j, r_{j'})$ is the travel time associated with mode $m$; \(r_j^{exit}\) denotes the designated exit point of building \(r_j\), and \(r_{j'}^{entry}\) denotes the designated entry point of building \(r_{j'}\). This conditional modeling ensures that if the classrooms are in the same building, only one indoor transit is considered, whereas if they are in different buildings, both the exit and entry indoor components are added to the outdoor transit time. To further calibrate these probabilities, we are preparing a survey for students and faculty to collect empirical data regarding their transportation preferences. The resulting data will be used to adjust the values of $p^{\text{out}}_m$ and $p^{\text{in}}_m$, thereby enhancing the accuracy and realism of the timetabling process. Figure~\ref{fig:highlights} provides an illustration of how spatial-temporal factors and multi-modal transportation options are integrated into the recommendation system.

\subsection{Scoring and Ranking Mechanisms}

The recommendation system evaluates and ranks potential course assignments using the composite scoring function defined in Equation~\ref{eq:scoring_function}. This equation—by integrating the normalized occupancy, travel distance, travel time, and travel floor scores—captures the trade-offs between efficient resource utilization and the ease of student transitions. In practice, for each feasible assignment \((c_i, r_j, t_k)\), the system computes the overall score \(S(c_i, r_j, t_k)\) based on the key metrics detailed in Section~\ref{sec:problem_statement}. 

Bottleneck courses—those with low scores due to excessive travel burdens or suboptimal classroom utilization—are then identified by comparing their individual scores to alternative assignments. By focusing on these critical courses, the system can explore alternative configurations and reassignments that yield improvements in the composite score \(Z\) as defined in Equation~\ref{eq:scoring_function}. Visual decision-support tools, such as radial plots and dynamic bar charts, further assist administrators in understanding the trade-offs among the different metrics, enabling informed and transparent decision-making.

Overall, the composite objective \(Z\) ensures that the final timetable achieves a balanced performance across all dimensions, aligning with both institutional priorities and user satisfaction.

\subsection{Iterative Optimization Process}

The iterative optimization process is the core mechanism for dynamically refining course timetables in response to real-time disruptions, evolving constraints, and diverse user preferences. Unlike static approaches that yield fixed schedules, this adaptive process continuously evolves through the following systematic steps:

\begin{figure}[!ht]
    \centering
    \includegraphics[width=\textwidth]{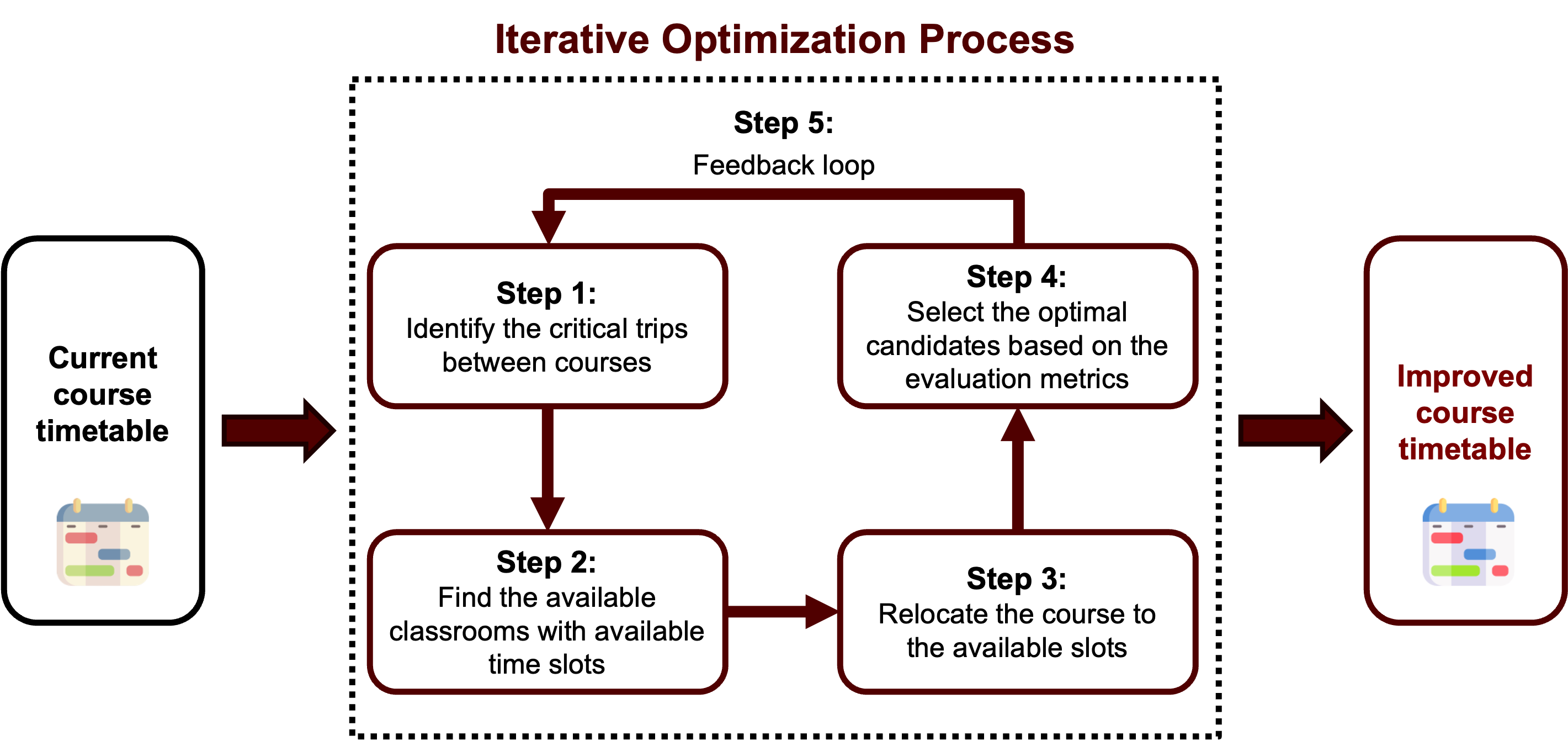}
    \caption{Overview of the Iterative Optimization Process.}
    \label{fig:process_overview}
\end{figure}

\paragraph{\textbf{Step 1: Identify Critical Issues}}  
The process begins with a thorough analysis of the current timetable to detect inefficiencies or constraint violations. For example, the system uses heatmaps to highlight areas where students experience excessive travel distances or where classroom capacities are overtaxed. This quantitative analysis provides immediate insight into which courses or transitions are problematic, allowing the system to prioritize areas for improvement.

\paragraph{\textbf{Step 2: Find Alternative Options}}  
Once critical issues are identified, the system explores feasible alternatives for the affected courses. This involves evaluating different classrooms and time slots that satisfy all hard constraints (e.g., capacity, travel time/distance, and floor transitions) while also considering soft constraints. The system employs dynamic bar charts and spatial visualizations to present the availability and proximity of alternatives, thereby enabling a comparative assessment of potential reassignments.

\begin{figure}[!t]
    \centering
    \includegraphics[width=\textwidth]{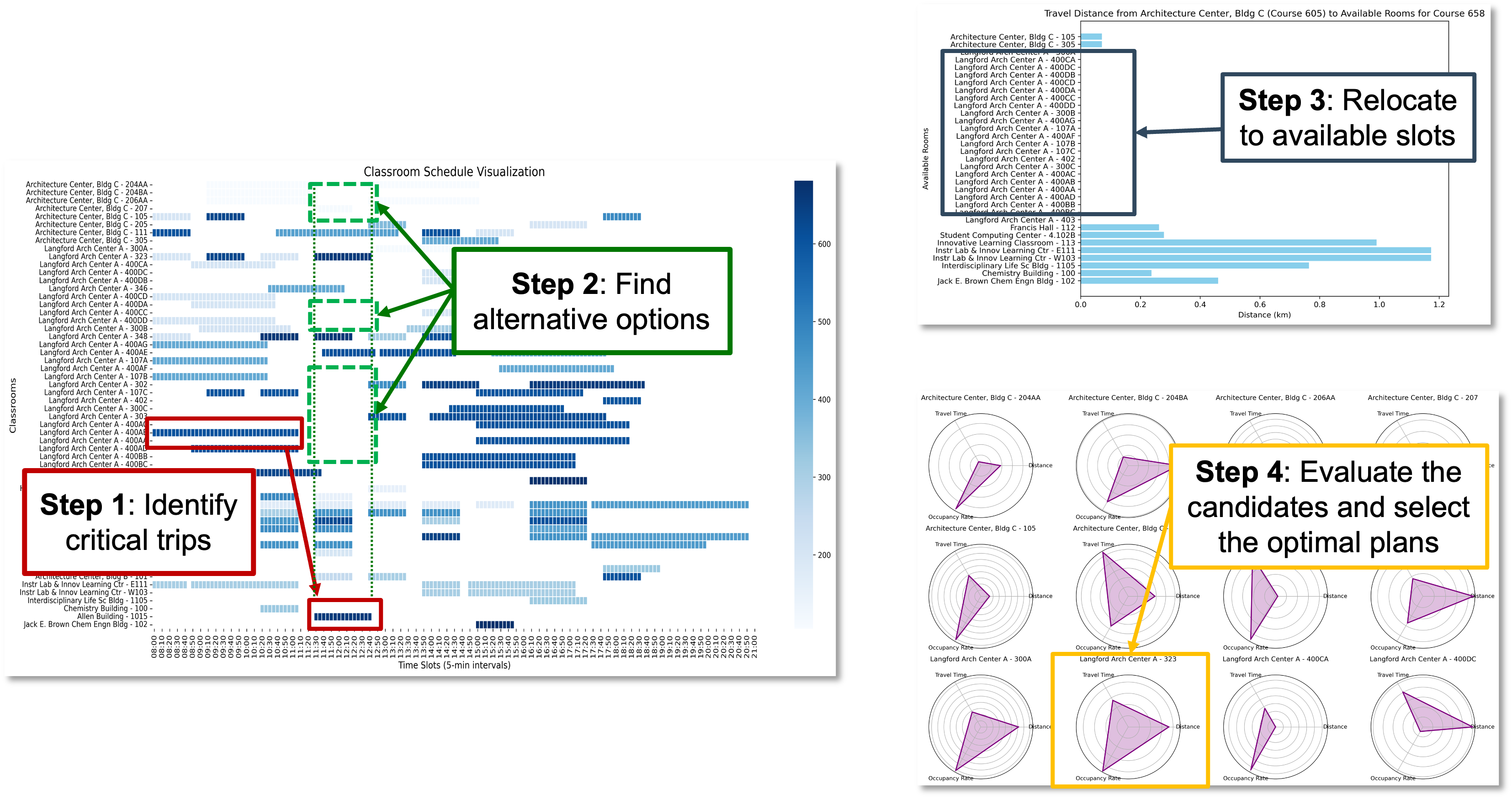}
    \caption{Detailed Steps of the Iterative Optimization Process.}
    \label{fig:process_steps}
\end{figure}

\paragraph{\textbf{Step 3: Relocate to Available Slots}}  
Based on the analysis, the system proposes reassignments for bottleneck courses. Each potential reallocation is rigorously evaluated using the composite scoring function (\ref{eq:scoring_function}). For instance, a course that is initially scheduled in a distant or overutilized room may be reassigned to a more accessible and better-matched classroom. Multiple candidate timetables are generated, each representing a different strategy for mitigating the identified issues.

\paragraph{\textbf{Step 4: Evaluate and Select Optimal Plans}}  
The candidate timetables are then ranked according to their composite scores. Visual tools such as radial plots allow administrators to compare the trade-offs across occupancy, travel distance, travel time, and floor transitions. The candidate with the highest overall score, indicating the best balance among the different metrics, is selected as the optimal plan. This rigorous evaluation ensures that improvements in one metric do not unduly compromise another.

\paragraph{\textbf{Step 5: Integrate Feedback and Iterate}}  
In the final step, the system incorporates feedback from stakeholders—including student and faculty satisfaction ratings and real-time operational updates—to further refine the timetable. Adjustments in the weights \(\alpha_1, \alpha_2, \alpha_3,\) and \(\alpha_4\) may be made based on this feedback, ensuring that the system remains aligned with institutional priorities and user needs. The iterative loop then recommences, continuously enhancing the timetable until a satisfactory and robust solution is achieved.

This iterative process, supported by dynamic visualizations and rigorous mathematical evaluation, not only improves timetable quality but also ensures transparency and adaptability. The methodical approach allows for continuous recalibration and fine-tuning, making it particularly effective in large, dynamic university environments where conditions and requirements may change frequently.

\section{Experiment Results} \label{sec:experiments}

\subsection{Data Integration and Visualization}

The initial phase of our experimental evaluation involved integrating multiple datasets related to campus facilities, classroom usage, and student movements, followed by a series of visual analyses to establish a baseline understanding of spatial-temporal dynamics. This integration ensured that data from different sources—such as building occupancy records, campus layout information, and real-time transition logs—could be combined in a unified format for subsequent optimization and recommendation tasks.

\begin{figure}[!ht]
    \centering
    \includegraphics[width=0.98\textwidth]{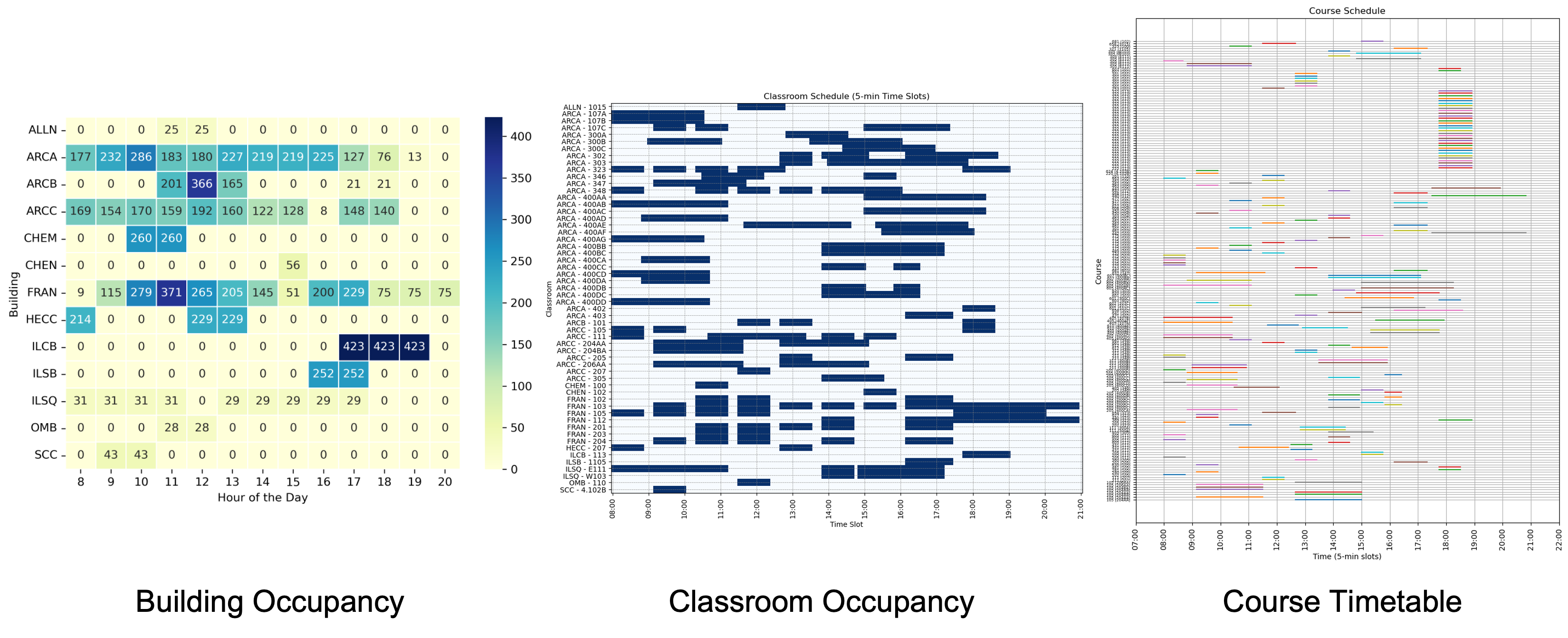}
    \caption{(a) Heatmap of hourly classroom occupancy for selected buildings, (b) Classroom schedules over time slots, and (c) Aggregated course schedules by day and time. These visualizations collectively highlight periods of high demand (midday hours) and potential underutilization in early mornings and late evenings.}
    \label{fig:data_viz}
\end{figure}

Figure~\ref{fig:data_viz} presents a heatmap illustrating hourly classroom occupancy across several campus buildings. Each cell indicates the number of students or classes occupying a specific building at a particular hour of the day, providing a clear picture of peak utilization times. Notably, the midday hours (approximately 10:00 AM to 2:00 PM) exhibit the highest occupancy, while early morning and late evening slots remain underutilized. This distribution aligns with common scheduling practices but also underscores the potential for more balanced allocations that could alleviate congestion and optimize room usage throughout the day.

Additionally, Figure~\ref{fig:data_viz} includes two supplementary views: one showing detailed classroom schedules over discrete time slots and another displaying aggregated course schedules across days. These perspectives help identify not only when buildings experience heavy demand, but also how classes are distributed among available time slots. Identifying clusters of heavily scheduled times or prolonged gaps in room usage can inform reassignments aimed at leveling out resource consumption.

\begin{figure}[!ht]
    \centering
    \includegraphics[width=0.98\textwidth]{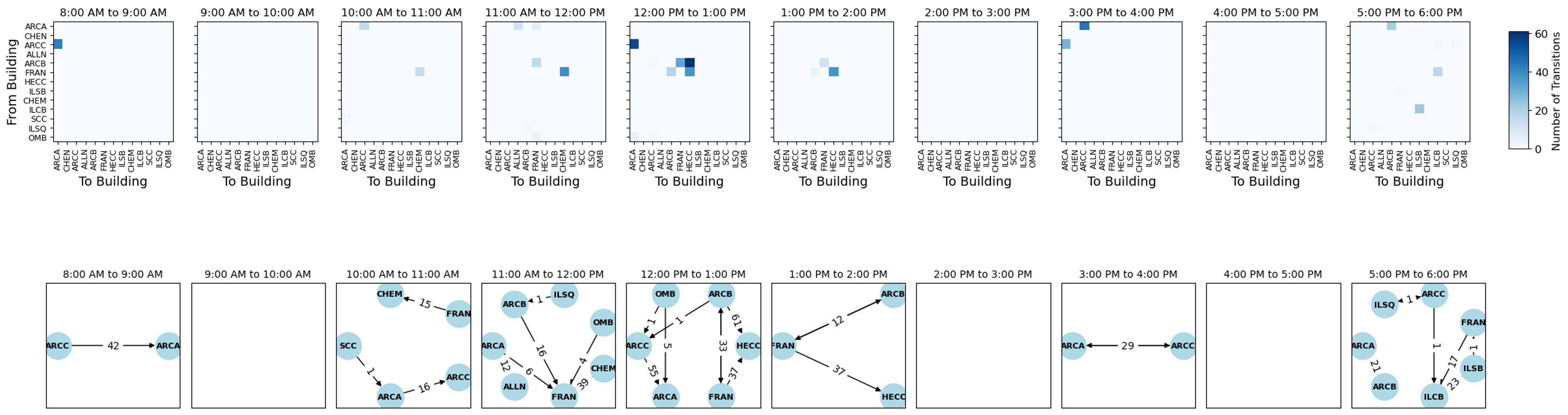}
    \caption{(Top) Transition matrices illustrating the number of student movements between buildings across different hours of the day, and (Bottom) corresponding network diagrams depicting inter-building travel flows. The size of each node or edge corresponds to the frequency of transitions.}
    \label{fig:transitions_viz}
\end{figure}

Figure~\ref{fig:transitions_viz} complements the occupancy analysis by depicting transitions between buildings over various hourly intervals. The top panel presents a series of heatmaps (or transition matrices), each representing a different time slice. Darker cells indicate higher volumes of student movements between specific building pairs, thereby revealing the spatial corridors with the greatest foot traffic. Meanwhile, the bottom panel displays network diagrams that visualize these inter-building flows. Larger nodes and thicker edges represent higher transition frequencies, highlighting building clusters—such as ARCC and FRAN—where travel is especially concentrated.

These integrated visualizations are critical for uncovering spatial-temporal inefficiencies. For example, frequent transitions between certain building pairs during peak hours suggest that students may face significant travel burdens, potentially leading to scheduling conflicts or lateness. Such insights motivate the need to reduce both travel distances and times within our composite scoring function (see Section~\ref{sec:problem_statement}). In addition, periods of underutilization revealed by the occupancy heatmap point to opportunities for distributing classes more evenly across the day, reducing congestion and improving resource allocation.

By synthesizing these data sources into intuitive visual formats, our framework not only identifies baseline patterns in classroom occupancy and student movements but also provides a robust foundation for targeted interventions. The subsequent stages of the experiment leverage these insights to guide the iterative optimization process, testing the effectiveness of reassignments and adjustments that aim to mitigate spatial-temporal bottlenecks and improve overall timetable quality.

\subsection{Bottleneck Identification}

Building upon the spatial and temporal insights uncovered during data visualization, the framework next focused on identifying \emph{bottleneck} courses—those that disproportionately exacerbate scheduling inefficiencies. In this context, inefficiencies manifest as either excessive travel distances between consecutive classes or prolonged travel times that disrupt student schedules. By scrutinizing both distance- and time-based metrics, the system pinpointed specific courses where reassignments could yield significant improvements in overall timetable quality.

\begin{figure}[!ht]
    \centering
    \includegraphics[width=0.95\textwidth]{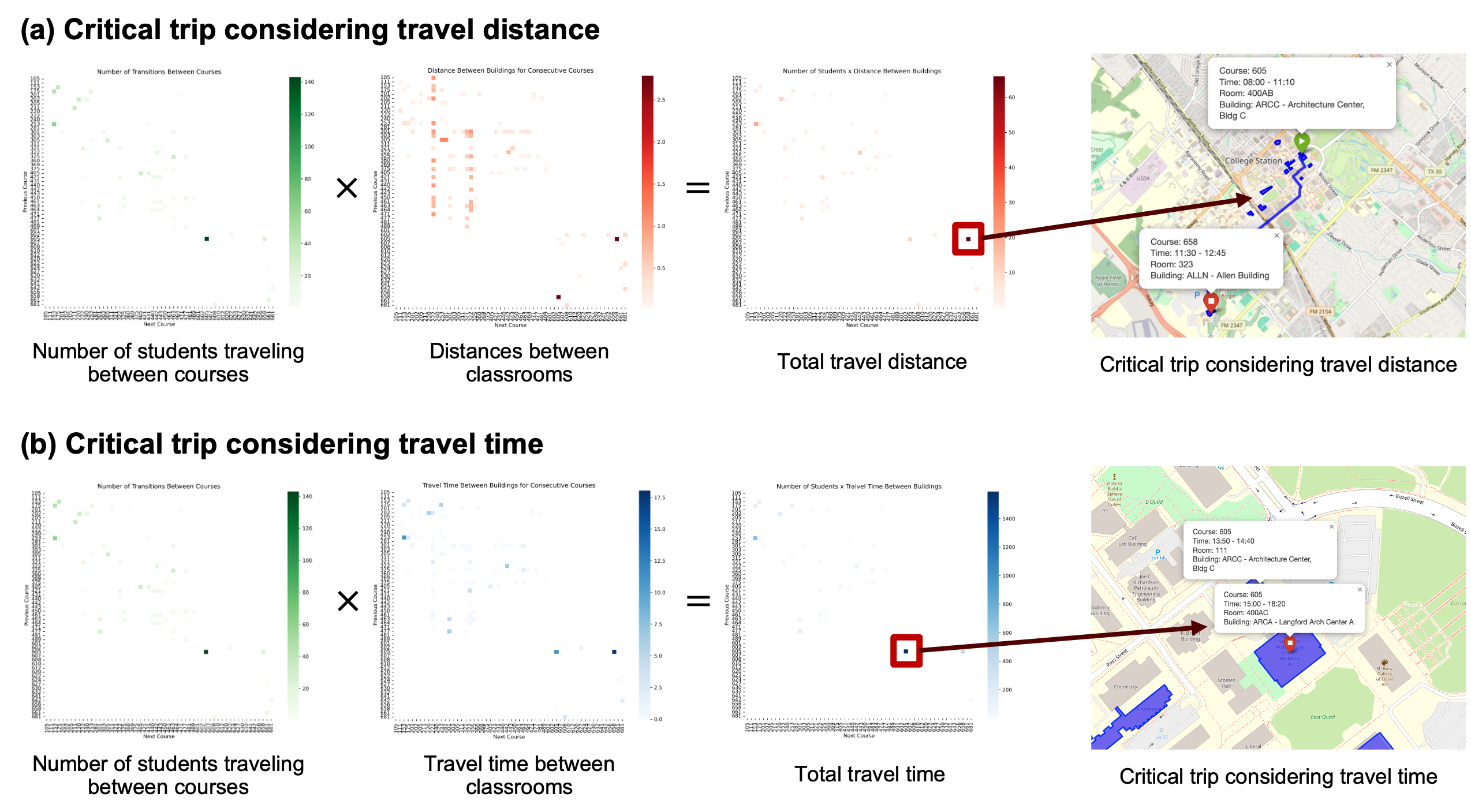}
    \caption{Illustrative views of bottleneck courses requiring excessive travel distances or times. The highlighted examples show transitions between peripheral buildings and central locations, as well as sequences with impractical travel durations. Pop-up annotations reveal the course details and specific paths contributing to these high-impact transitions.}
    \label{fig:critical_trip}
\end{figure}

Figure~\ref{fig:critical_trip} presents a two-tiered analysis of these bottlenecks. The top row highlights courses associated with long travel distances, frequently involving outlying buildings in the west campus. These locations often require students to traverse considerable portions of campus, which is especially problematic when classes are scheduled back-to-back without sufficient transition time. The bottom row focuses on instances where cumulative travel time over all the students exceed acceptable thresholds, in some cases magnified by vertical transitions via staircases or elevators.

In more concrete terms, certain course sequences were flagged for requiring nearly 20 minutes of travel between buildings—an impractical interval that jeopardizes punctuality and adds stress for both students and faculty. As an example, transitions between Architecture Center, Building C (ARCC) and the Allen Building (ALEN) emerged as a critical trip due to a combination of distance and reliance on congested walkways. These findings underscore the necessity of an adaptive scheduling approach—one that actively accounts for spatial feasibility and user convenience. By dynamically identifying and targeting these high-impact bottlenecks, the system lays the groundwork for iterative reallocations, aligning with the composite scoring and iterative optimization strategies outlined in Section~\ref{sec:methodology}.

\subsection{Metric-Based Evaluation}

Having identified the most problematic bottlenecks, the framework next produced and evaluated alternative scheduling solutions using the composite scoring function detailed in Section~\ref{sec:methodology}. This scoring mechanism incorporates metrics such as travel distance, travel time, occupancy improvements, and floor transitions—each weighted to reflect institutional priorities. By measuring both the spatial and temporal feasibility of course assignments, the system provides a comprehensive view of timetable quality, ensuring that remedial actions target critical inefficiencies without introducing new conflicts.

\begin{figure}[!ht]
    \centering
    \includegraphics[width=0.95\textwidth]{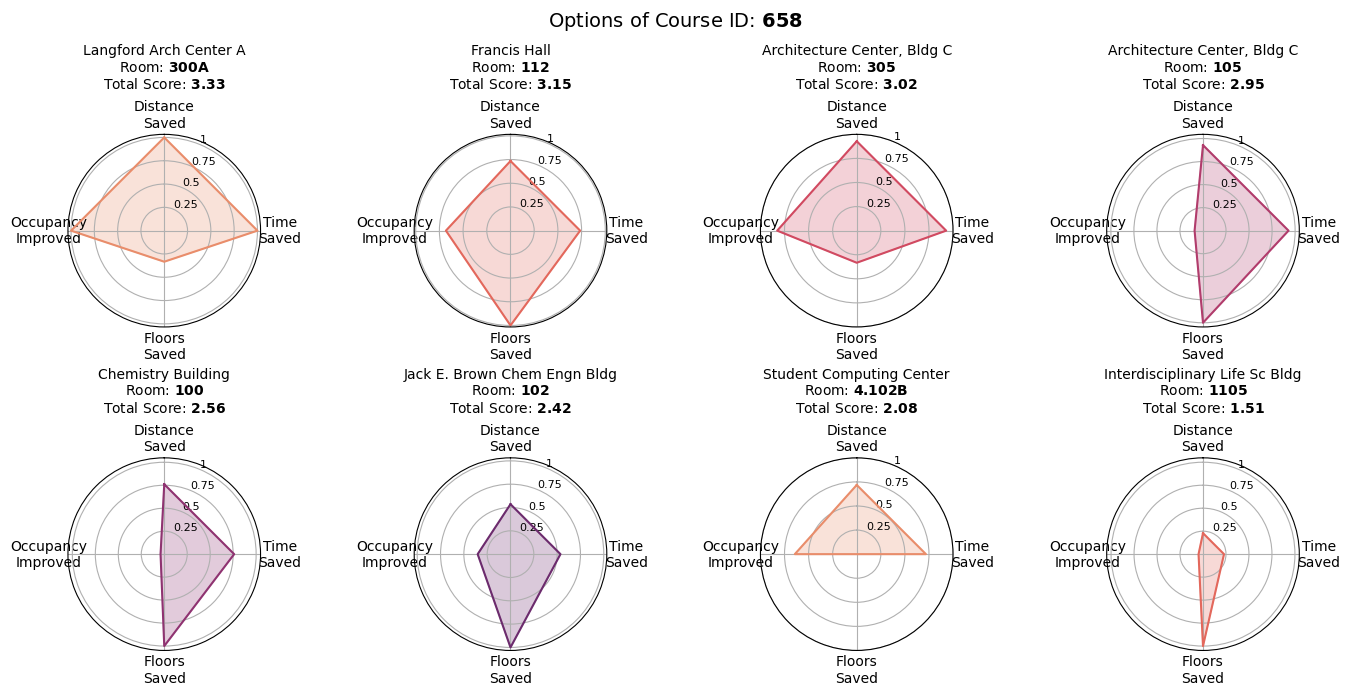}
    \caption{Radar plots of alternative assignments for Course ID 658, focusing on minimizing travel distance while maintaining or improving occupancy. Each radial axis corresponds to a different metric in the composite scoring function, with the overall score displayed above each plot.}
    \label{fig:eval_dist}
\end{figure}

Figure~\ref{fig:eval_dist} presents a selection of potential classroom reassignments for Course ID 658, primarily aimed at reducing travel distance. Each radar plot represents a unique scenario, where the radial axes indicate how each metric—distance, time, occupancy, and floor transitions—is affected. For instance, moving the course to \emph{Architecture Center, Building C} (Room 105) can significantly reduce travel distance and time, as well as the number of floor changes, but may offer only moderate occupancy gains. In contrast, placing the course in \emph{Francis Hall} (Room 112) may yield a more balanced outcome across several metrics, leading to a higher overall score. Meanwhile, reassigning it to \emph{Langford Arch Center A} (Room 300A) could maximize time savings at the expense of additional floor transitions. By examining these trade-offs—such as weighing a 0.4 boost in occupancy against a 0.15 decrease in distance savings—decision-makers can select the option that best serves institutional objectives.

\begin{figure}[!ht]
    \centering
    \includegraphics[width=0.95\textwidth]{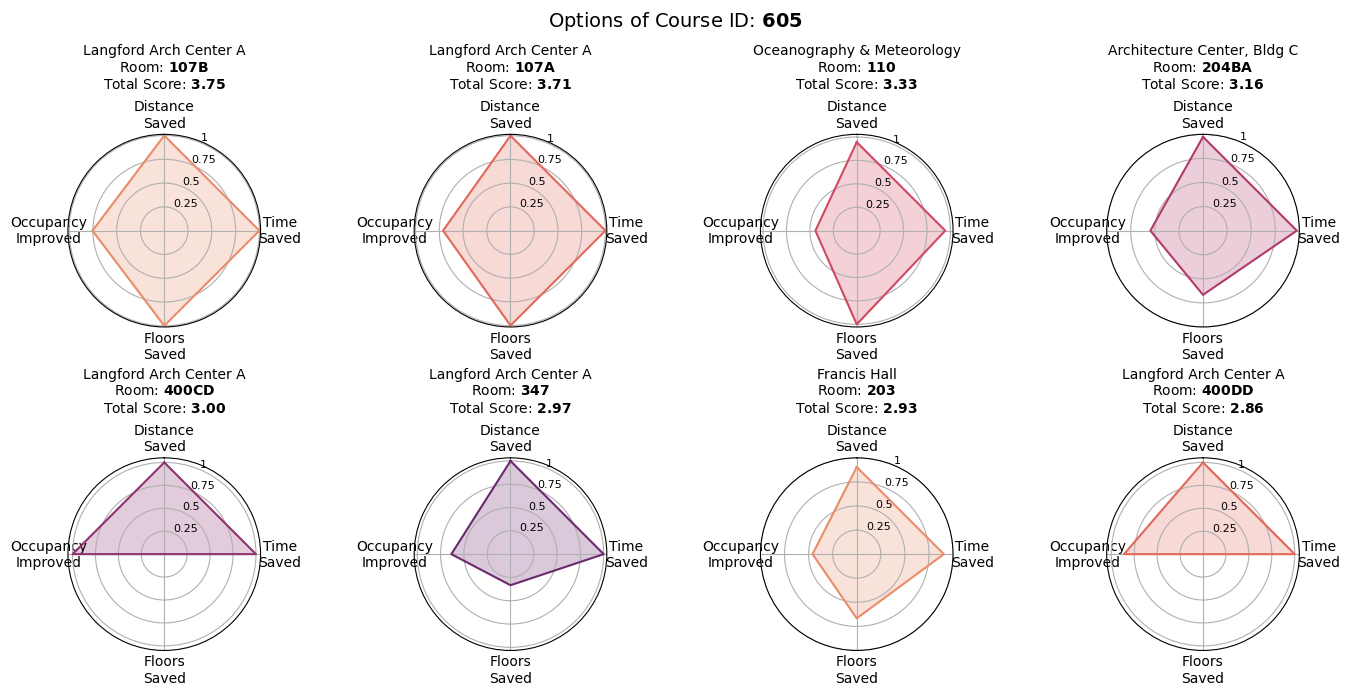}
    \caption{Radar plots of alternative assignments for Course ID 605, focusing on minimizing travel time while balancing other metrics in the composite scoring function.}
    \label{fig:eval_time}
\end{figure}

Figure~\ref{fig:eval_time} similarly illustrates scheduling alternatives for Course ID 605, where minimizing travel time is the principal objective. For example, assigning the course to \emph{Oceanography \& Meteorology Building} (Room 110) might notably reduce floor transitions but marginally lower occupancy. Conversely, relocating it to \emph{Langford Arch Center A} (Room 107B) could produce a more balanced outcome across multiple metrics, resulting in the highest composite score. 

Overall, these metric-based assessments demonstrate the system’s capacity to generate practical, data-driven recommendations that support the twin goals of flexibility and user-centered optimization. By quantifying changes in travel distance, travel time, occupancy, and floor transitions, the framework ensures that proposed reassignments systematically address previously identified spatial-temporal bottlenecks, thereby improving the robustness and efficiency of the course timetable.

\subsection{Comparison of Original and Optimized Schedules}

\begin{figure}[!ht]
    \centering
    \includegraphics[width=1.0\textwidth]{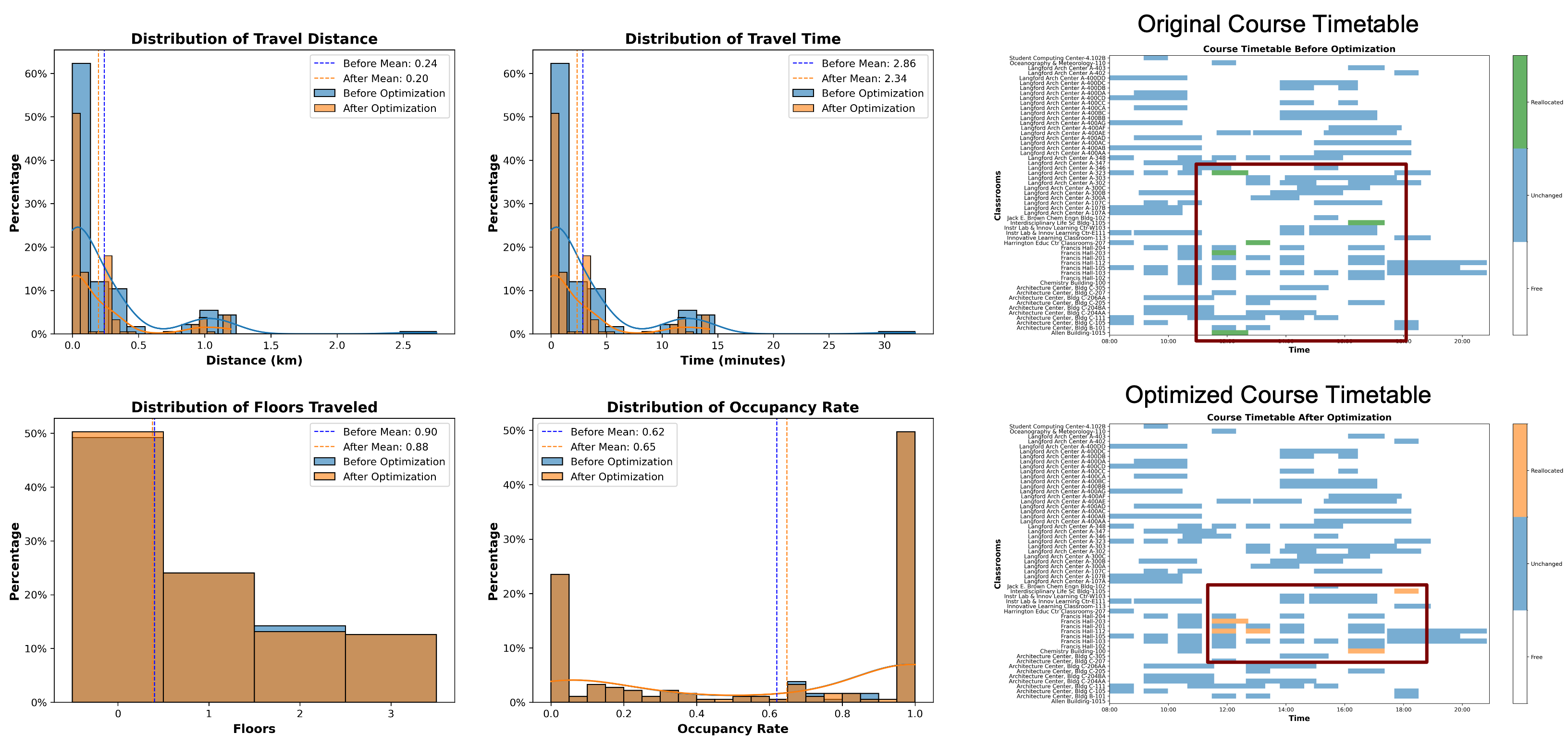}
    \caption{Distribution of travel distance, travel time, floors traveled, and occupancy rate for the original vs.\ optimized schedules, as well as a heatmap comparison of the course timetable. Vertical dashed lines in the histograms indicate mean values.}
    \label{fig:optimization_results}
\end{figure}

To further assess the impact of our iterative recommendation framework, we compared the original course schedule with the optimized schedule after a set number of iterations, focusing on the four core metrics discussed in Section~\ref{sec:problem_statement}. Figure~\ref{fig:optimization_results} illustrates the distribution of travel distance, travel time, number of floors traveled, and occupancy rate both before and after optimization, along with a side-by-side view of the original and updated timetables.

In the histograms on the left, the optimized schedule shows a pronounced shift toward lower travel distances and shorter travel times. The reduction in outliers—students or faculty facing especially long commutes—highlights the system’s effectiveness in targeting spatial-temporal bottlenecks identified during earlier analyses. Additionally, the histogram of floors traveled reveals fewer high-floor transitions in the optimized schedule, suggesting that reassignments have reduced vertical movements between classes. Meanwhile, the occupancy rate distribution indicates a slightly more balanced utilization of classrooms.

On the right, the timetable comparison visually underscores these improvements. The highlighted regions in the optimized schedule reveal reassignments that alleviate congested corridors and better align class locations with student flow. Courses that once required excessive travel have been relocated to buildings with closer proximity or fewer vertical transitions, and some underutilized slots have been repurposed to reduce midday congestion.

Collectively, these results demonstrate how iterative feedback loops—guided by the composite scoring function—can systematically refine scheduling decisions. By quantifying improvements in travel distance, travel time, floors traveled, and occupancy, the framework ensures that adjustments made in each iteration yield tangible benefits. The final outcome is a more user-centric timetable that balances resource utilization with spatial-temporal feasibility, reflecting the adaptive nature of the proposed methodology.

\section{Discussion} \label{sec:discussion}

The experimental results highlight the effectiveness of our recommendation-based framework in addressing the intricacies of large-scale course timetabling. Under standard conditions, the system produced schedules comparable to those generated by conventional optimization techniques, while its true advantages emerged when confronted with rapidly changing constraints. In scenarios such as peak enrollment surges or unforeseen building closures, the framework adaptively reallocated courses to maintain high classroom utilization, demonstrating resilience and responsiveness beyond the capabilities of static models.

A key insight from these experiments was the reduction in travel-related inefficiencies for students and faculty. By incorporating spatial-temporal data, the framework pinpointed and remedied courses that contributed disproportionately to lengthy transitions between classes. In practical terms, this meant fewer instances of extended walks or rushed travel times, which in turn fostered an improved sense of schedule convenience among users. Furthermore, the system’s iterative approach enabled it to react quickly to unexpected events by refining course assignments incrementally, rather than recalculating the entire timetable. Another notable outcome was the framework’s capacity to optimize classroom utilization while respecting critical constraints such as room capacity and scheduling conflicts. By reallocating courses to underused spaces during busy periods, the system minimized congestion without requiring a full overhaul of existing assignments. This flexibility illustrates the framework’s user-centric ethos, wherein data-driven insights guide reassignments that benefit both institutional objectives and individual needs.

The adaptive, feedback-driven mechanism underlying the framework also underscores its potential for broader applications. Its ability to integrate real-time updates and respond to evolving conditions makes it well-suited for dynamic environments beyond academic scheduling, such as urban resource management or facility planning. Nonetheless, the system’s reliance on high-quality, comprehensive data remains an essential consideration; any inconsistencies or gaps in input data could diminish its overall effectiveness.

Overall, these findings suggest that the proposed recommendation-based framework offers a robust and scalable solution to the complex challenge of course timetabling. By coupling spatial-temporal analytics with iterative refinement and user-centric design, the system not only enhances operational efficiency but also aligns scheduling decisions with the evolving needs of a diverse campus community.

\section{Conclusion} \label{sec:conclusion}

This study introduced an adaptive, recommendation-based framework for course timetabling underpinned by the Texas A\&M Campus Digital Twin, with a central emphasis on \emph{human mobility}. Moving beyond one-shot, static scheduling, the proposed approach reframes the timetable as a dynamic mobility-shaping policy that must remain feasible and useful as constraints, preferences, and operational conditions evolve. In this view, each assignment of a course to a room and time slot is also a decision about movement: it structures when and where students and instructors travel, how they interact with indoor-outdoor infrastructure, and how mobility burdens accumulate across a day. By combining collaborative and content-based filtering with an iterative feedback loop, the framework continuously generates and refines ranked scheduling alternatives, enabling decision-makers to compare mobility-resource trade-offs rather than relying on a single rigid solution. The digital twin serves as the computational substrate that turns mobility from an implicit side effect into an explicit, measurable objective through realistic spatial-temporal costs, supporting responsive scheduling while preserving transparency in how mobility outcomes and institutional constraints are balanced.

Experimental results demonstrate that the framework can systematically reduce mobility friction, lowering travel distance, travel time, and vertical transition burdens, while improving classroom utilization and overall timetable quality. Through bottleneck-oriented refinement, the system identifies courses and transitions that disproportionately amplify movement costs (e.g., long cross-campus transfers, repeated vertical changes, or tightly constrained back-to-back walks) and targets them for localized adjustments, avoiding unnecessary global reshuffling. The resulting schedules exhibit fewer extreme outliers in travel burden, improved feasibility and reliability of consecutive-class transitions, and more balanced occupancy patterns across rooms and time windows. Importantly, these gains are achieved while respecting hard requirements such as capacity and conflict avoidance, showing that mobility-centered objectives and resource efficiency can be optimized within a unified decision process. Beyond quantitative improvements, the framework provides a practical pathway for administrators to operationalize human-centered scheduling by making mobility impacts auditable and comparable across candidate plans.

Despite these strengths, the framework relies on high-quality, comprehensive digital twin data, underscoring the importance of robust integration and validation. Future work can extend the framework with group-wise human movement and interaction-aware relational reasoning to optimize collective congestion dynamics, not only individual travel costs~\cite{zhang2024spatial, wu2025hypergraph}. LLM-enabled explainability can translate mobility analytics into stakeholder-facing rationales for proposed changes~\cite{liang2024exploring, gong2024mobility, wu2025v2x}, while agentic scheduling systems can coordinate specialized agents (mobility, capacity, accessibility, policy) to negotiate and verify timetable updates within the digital twin~\cite{du2025cams, zhang2024agentic}. Finally, evolving toward a closed-loop ``sense--predict--simulate--act'' digital twin pipeline with real-time sensing and predictive models can enable proactive, uncertainty-aware scheduling that incorporates broader objectives such as equity, safety, and sustainability~\cite{fan2022integrating,irfan2024toward, wu2025digital}.


\bibliographystyle{ieeetr}
\bibliography{ref}

\end{document}